\documentstyle[12pt]{article}
\input{epsf}

\newcommand{\newsection}{                     % Numeration of eqs. is automatic
                         \setcounter{equation}{0}\section}
\renewcommand{\appendix}[2]{\bigskip {\subsection*{Appendix #1. \\ #2}}}
\newcommand{\semi}{{\\ \hs}}
\newcommand{\rf}[1]{(\ref{#1})}
\newcommand{\eq}[1]{Eq.~(\ref{#1})}

\newcommand{\beq}{\begin{equation}}
\newcommand{\eeq}{\end{equation}}
\newcommand{\bea}{\begin{eqnarray}}
\newcommand{\eea}{\end{eqnarray}}

\renewcommand{\l}{\lambda}
\newcommand{\om}{\omega}
\newcommand{\Om}{\Omega}
\newcommand{\oh}{{1\over 2}}
\newcommand{\ud}{{1\over 2}}
\newcommand{\Tr}{{\,\rm Tr}\:}
\newcommand{\tr}{{\,\rm tr}\:}

\newcommand{\e}{{\,\rm e}\,}

\newcommand{\hs}{\hspace{0.7cm}}

\newcommand\ee[1]{{\rm e}^{^{\textstyle#1}}}
\renewcommand\d[1]{{\rm d}\!#1\,}
\renewcommand\e{{\rm e}}
\renewcommand\Re{{\rm Re\,}}
\renewcommand\Im{{\rm Im\,}}
\newcommand\moy[1]{{\left< #1 \right>}}
\newcommand\bra[1]{{< #1 |}}
\newcommand\ket[1]{{| #1 >}}
\newcommand\roc{{\cal O}}
\renewcommand\O{{\cal O}}
\newcommand\s[1]{\sqrt{(#1-a)(#1-b)}}
\renewcommand\t[1]{\tilde{#1}}

\begin{document}
\topmargin 0pt
\oddsidemargin 5mm
\headheight 0pt
\headsep 0pt
\topskip 9mm

{\flushright{\vbox{\baselineskip12pt\hbox{SPhT/97-031}}}}
%{Large Random Matrices: Eigenvalue Distribution}
{\vskip 2cm \begin{center} \bf \Large Eigenvalue distribution of large random matrices,\\
from one matrix to several coupled matrices. \end{center}}

\vskip 1.5cm
\centerline{B. Eynard}

\bigskip{\baselineskip14pt
\centerline{Service de Physique Th\'eorique de Saclay\footnote{Laboratoire de la Direction des Sciences de la Mati\`ere du Commissariat \`a l'Energie Atomique}}
\centerline{F-91191 Gif-sur-Yvette Cedex, FRANCE}
\centerline{Email: eynard@wasa.saclay.cea.fr}}
\bigskip

\vskip 1cm
%
%---------------------------------------------------------------------
%
%                          ABSTRACT
%
%
\centerline{\bf Abstract:}
It has been observed that the statistical distribution of the eigenvalues of random matrices possesses universal properties, independent of the probability law of the stochastic matrix.
In this article we find the correlation functions of this distribution in two classes of random hermitian matrix models: the one-matrix model, and the two-matrix model, although it seems that the methods and conclusions presented here will allow generalization to other multi-matrix models such as the chain of matrices, or the $O(n)$ model.
We recover the universality of the two point function in two regimes: short distance regime when the two eigenvalues are separated by a small number of other eigenvalues, and on the other hand the long range regime, when the two eigenvalues are far away in the spectrum, in this regime we have to smooth the short scale oscillations.
We also discuss the universality properties of more than two eigenvalues correlation functions.


\vskip 1cm


%
%
%---------------------------------------------------------------------
%

\vskip.3in
{\flushleft{03/97, for Nuclear Physics B (FS)}}

\eject

It has been observed experimentally \cite{rCG,rHWB} for disordered systems that the distribution of the energy levels of such systems is universal in some regime.
For instance the connected correlation function between two levels separated by a small number of other levels does not depend on the system, while the density of levels is very dependent of the specific details of the system.
Such a phenomenon is also observed in numerical simulations of large random matrices \cite{rCamarda}.
The consequence of this is that the correlation functions can be obtained from a gaussian model.
This conjecture has been proved in the special case of a one hermitian matrix model \cite{rBrZe,rWigner}.
It has also been noted that the connected correlation functions of more than two eigenvalues should present also a stronger universality than the density itself.
Br\'ezin and Zee \cite{rBrZe} have calculated explicitly correlation functions of eigenvalues of a class of stochastic hermitian matrices of large size $N$.
They have found that some statistical properties of the eigenvalues are
universal in the large $N$ limit, and can thus be obtained from  the
correlation functions of the gaussian model. 
In addition, they have discovered that the two-point correlation function,
after smoothing of the short scale oscillations, is universal while all other
smoothed correlations vanish at the same order.
This property could be understood by a renormalization group analysis
\cite{rBrZi,rHINS} which indicates that the gaussian model is a stable
fixed point in the large $N$ limit.

Random matrix models appear also in another field of physics: they have been discovered to describe a regularized theory of random surfaces and string theory \cite{rBBIPZ,rDSBKGM}.
The string theoricians have thus developed very powerful tools for the study of the critical properties of these models, as orthogonal polynomial methods, saddle point approximations or loop equations.

The analysis of \cite{rBrZe} is based on the, by now standard method of orthogonal
polynomials.
An essential ingredient in the final answer is a proposed ansatz for an asymptotic form of the orthogonal polynomials $P_n$ in the limit
$N\to\infty$ and $N-n$ finite.
In \cite{rBrZe} the ansatz is verified in the case of even integrands, and only up to an unknown function. 
They also examine the smoothed correlation functions in the large scale regime, using a diagrammatic method.
\par
Here, we propose a direct proof of the ansatz for the orthogonal polynomial, using a saddle point method, which does not depend on the parity of the integrand,  and which allows to determine the previously unknown function.
Moreover, we extend this method to another class of matrix-models known in 2D gravity as the two-matrix model.
This model presents a larger class of critical points than the one-matrix-model, it is therefore interresting to find if it presents the same universality.

In addition, using the loop equations approach \cite{rAJM,rACKM},
we can generalize the diagrammatic method of \cite{rBrZe} to find all the connected
correlation functions of the one-matrix-model, and discuss their universality properties.
This recursive method allows explicit calculations, and we write the three and four point function.
This method could also be generalized to another class of matrix models: the $O(n)$ model, which has the same kind of loop equations and topologial expansion procedure \cite{rEyKr}.
\newsection{The 1-Matrix-model}
Let us first explain the problem and recall the method used in ref.~\cite{rBrZe}
to  explicitly evaluate the eigenvalue correlation functions.\par

We consider a $N\times N$ stochastic hermitian matrix $M$  with a probability
law of the form: 
\beq\label{eprob}
{\cal P}(M)={1\over Z}\, \ee{-N\tr V(M)},
\eeq
where $V(M)$ is a polynomial, and $Z$ the normalization (i.e.\ the partition
function).
We want to derive the asymptotic form for $N$ large of various eigenvalue correlation functions.
All of them can be obtained from the correlation functions of the operator
$\O(\lambda)$:
\beq\label{eOp}
\O(\l)={1\over N}\tr \delta(\l-M)={1\over N}\sum_{i=1}^N 
\delta\left(\l-\mu_i\right)
\eeq
(the $\mu_i$ being the eigenvalues of $M$).
Indeed, for any set of functions $(f_1,...,f_k)$, one has:
$${1\over N^k}\left< \tr f_1(M)\ldots\tr f_k(M)\right>=\int \d\!\l_1 f_1(\l_1)\ldots
\d\!\l_k f_k(\l_k)\, \left< \O(\l_1)\ldots \O(\l_k)\right>.$$
The correlation functions of the operator $\O(\l)$ can in turn be expressed in terms of the partially integrated eigenvalue distributions like $\rho(\l)$ the density of eigenvalues, which is the probability that $\l$ belongs to the spectrum of $M$, $\rho_2(\l,\mu)$ the probability that $\l$ and $\mu$ are simultaneously  eigenvalues of $M$ and more generally $\rho_n(\l_1,\ldots,\l_n)$ the probability that $\l_1$ ...$\l_n$ are simultaneously eigenvalues of $M$.
For $\lambda_1\ne \lambda_2 \ldots \ne \lambda_n$ (else some additional contact terms have to be added)
we find
$$\left<\O(\lambda_1)\O(\lambda_2)\ldots \O(\lambda_n)\right>= 
{1\over N^n}{N!\over (N-n)!} \rho_n(\l_1,\ldots,\l_n).$$

Actually the interesting functions are not directly the $\O(\lambda)$
correlation functions, but their connected parts.
Indeed, at leading order, when $N\to\infty$, we have the factorization property
$$\left<\O(\lambda_1)\O(\lambda_2)\ldots {\cal O}(\lambda_n)\right> \sim \prod_{i=1,\ldots, n} \left<{\cal O}(\lambda_i)\right>,$$ 
and thus no new information can be obtained from the complete $n$-point function.
The connected function which will be denoted
$$\left< \O(\l_1)\ldots \O(\l_n) \right>_{\rm conn}=\roc_n(\l_1,\ldots,\l_n) $$
The method of orthogonal polynomials allows to determine directly all these connected functions from only one auxiliary kernel $\kappa(\lambda,\mu)=\kappa(\mu,\lambda)$ (see refs.~\cite{rDyson,rBrZe,rMehta} or appendix~2 for details).
For instance we have:
$$\begin{array}{rl}
 \roc_1(\l) & = \rho(\l)=\kappa(\l,\l) ,\cr
\roc_2(\l,\mu) & = -\kappa^2(\l,\mu) ,\cr
\roc_3(\l_1,\l_2,\l_3) & =2\kappa(\l_1,\l_2)\kappa(\l_2,\l_3)
\kappa(\l_3,\l_1),\cr
\roc_4(\l_1,\l_2,\l_3,\l_4) & =-2\kappa(\l_1,\l_2)\kappa(\l_2,\l_3)
\kappa(\l_3,\l_4)\kappa(\l_4,\l_1)
\cr & \quad -2\kappa(\l_1,\l_3)\kappa(\l_3,\l_2)\kappa(\l_2,\l_4)
\kappa(\l_4,\l_1) \cr
 & \quad -2\kappa(\l_1,\l_2)\kappa(\l_2,\l_4)\kappa(\l_4,\l_3)
\kappa(\l_3,\l_1),\cr
\end{array}$$
and analogous expressions for larger values of $n$:
\beq\label{erocnkappa}
 \roc_n=(-1)^{n+1}
\sum_{{\rm cyclic\, permutations}\, \sigma\,}
\,\prod_{i=1}^n \kappa\left(\l_i,\lambda_{\sigma_i}\right) 
\eeq
The kernel $\kappa$ is simply related to the polynomials $P_n$ orthogonal with respect to the measure $\d\lambda\, \ee{-NV(\lambda)}$:
\beq\label{eKdef}
\kappa(\l,\mu) ={1\over N}\sum_{n=0}^{N-1} P_n(\l)P_n(\mu)\exp[-(N/2)(V(\l)+V(\mu))]
\eeq
With the help of the Darboux-Christoffel theorem (Appendix~3) we can rewrite it:
\beq\label{eKDC}
\kappa(\l,\mu) \propto {1\over N}
{P_N(\l)P_{N-1}(\mu)-P_{N-1}(\l)P_N(\mu)\over \l-\mu}
\exp[-(N/2)(V(\l)+V(\mu))] .
\eeq
So, the asymptotic evaluation of correlation functions is reduced to an evaluation of the kernel $\kappa(\l,\mu)$ and thus of the orthogonal polynomials $P_n(\l)$ in the peculiar limit: $N$ large, $N-n$ finite, and $\l\in[a,b]$~(~$[a,b]$ being the support of $\rho(\l)$~).
 
The ansatz proposed in \cite{rBrZe} in the case of even potentials $V(M)$ (for
which $b=-a$)
was: 
$$ P_n(\l)\, \propto\, \ee{N{V(\l)/ 2}}{1\over \sqrt{f(\l)}}\cos{\bigl(
N\zeta(\l)+(N-n)\varphi(\l)+\chi(\l)\bigr)},$$ 
where
$$\begin{array}{rl}
\l & = a\cos\varphi\,, \cr
f(\l) & =a\sin\varphi\,, \cr
{\d\over \d\l}\zeta(\l) & =-\pi\rho(\l),\cr
\end{array}$$
$\chi(\l)$ remaining undetermined, except in the gaussian and quartic
cases, for which: $\chi(\l)=\varphi/2 -\pi/4$. 

We will prove below that this ansatz is still true for any $V$, and that
$\chi$ is always of the form $\chi=\varphi/2+{\rm const}$. The method 
which leads to the proof is also interesting in itself because it uses some
general tools of the saddle point calculations of the one-matrix model
\cite{rBBIPZ,rDSBKGM}.
This method will be extended to the multi-matrix model in the next section.
\subsection{Orthogonal Polynomials}

We thus have to consider the set of polynomials ${P_n}$ ($n$ is the degree),
orthogonal with respect to the following scalar product: 
\beq\label{ePns}
\left<P_n \cdot P_m\right> = \delta_{nm} = \int \d\l\, \ee{-N
V(\l)}\, P_n(\l)P_m(\l) .
\eeq
Remarkably enough an explicit expression of these
orthogonal polynomials in terms of a hermitian matrix integral can be derived
(see appendix~1, or \cite{rSze}): 
\beq\label{ePnexact}
P_n(\l)= \sqrt{n+1\over Z_n Z_{n+1}}\int {\rm d}^{n\times n}\! M\, \ee{-N \tr V(M)}\, \det(\l -M),
\eeq
where $M$ is here a $n\times n$ hermitian matrix, and 
$Z_{n}$ the partition function \rf{eprob}\ corresponding to
$n\times n$ matrices. We will use this expression 
to evaluate $P_n$ in the relevant limit, i.e\ $N\gg 1$ and $N-n=O(1)$ 
by the steepest descent method.
Let us first consider the matrix integral:
$$ Z(g,h,\l)=\int d^{n\times n}\! M\,\, \ee{-(n/g) \tr {\cal V}(M)}, $$
$$\qquad {\cal V}(z)= V(z) -h\ln{(\l-z)}. $$
With these definitions $P_n(\l)$ is proportional to $Z(g=n/N,h=1/N,\l)$, and thus
we need $Z$ or equivalently the free energy $F_n=-n^{-2}\ln Z$ for $h$ and
$g-1$ small.  

As we are interested only in the $\l$ dependance of $F$, let us differentiate
$F$ with respect to $\l$: 
$$ {\partial F\over \partial \l}=-{h\over g} \om(z=\l,g,h,\l)$$
where $\om(z)$ is the resolvent:
\beq\label{eresolv}
\om(z)={1\over n}\moy{ \tr{1\over z-M}} .
\eeq

Since we want the asymptotic expression of $ \om(\l)n^2 h/g$, we
need $\om$ up to the order $1/n$.
It is known from the random matrix theory \cite{rDGZ}\ , that $F_n$, and also $\om$ have an expansion in powers of $1/n^2$ which is the topological expansion.
At order $1/n$, only the contribution of the sphere is required.
We thus replace $\om$ by its dominant contribution obtained by the saddle
point method. 
\medskip
With this approximation, $\om$ may be written:
$$\om(z)={1\over n}\sum_{i=1,\ldots, n} {1\over z-\l_i}$$
where the $\l_i$ are the eigenvalues verifying the saddle point equation:

$${1\over g}{\cal V}'(\l_i)={2\over n}\sum_{j\neq i} {1\over \l_i-\l_j}\,,$$
i.e. they extremize the integrand of 
$$ Z(g,h,\l)\propto \int \prod_{i=1}^{n} \ee{-N{\cal V}(\l_i)}d\!\l_i\,
 \prod_{i<j} (\l_i-\l_j)^2 \,.$$
In the large $n$ limit the $\l_i$ are distributed along an interval $[a,b]$
with a continuous density $\rho(\l)$ \cite{rWigner,rMehta}. Then:
\beq\label{eomrho}
\om(z)=\int_a^b \d \mu\, \rho(\mu) {1\over z-\mu}
\eeq
and the saddle point equation becomes:
\beq\label{elinear}
{1\over g}{\cal V}'(\mu)=\om(\mu+i0)+\om(\mu-i0) \qquad {\rm for}\quad\mu\in [a,b]
\eeq
Note an important property of this equation: At $a$ and $b$ fixed it is linear
and therefore the derivatives of $\om$ with respect to $g$ or $h$ will 
also satisfy a linear equation.

At leading order  we introduce the resolvent $\om_0(z)=\om(z,g=1,h=0,\l)$, and
write: 
$$ g\om(z,g,h,\l)=\om_0(z)+(g-1)\Om_g(z,\l)+ h \Om_h(z,\l) + O(1/n^2),$$
where we have defined the two functions:
$$\Om_g=\left. {\partial g\om\over \partial g}\right|_{g=1,h=0},\qquad
\Om_h=\left. {\partial g\om \over \partial h}\right|_{g=1,h=0}.$$

The function $\om_0(z)$ is the resolvent of the usual one-matrix model, and
from eqs.~\rf{eomrho}\ and \rf{elinear}\ we obtain:
$$\om_0(z\pm i0)={1\over 2} V'(z) \mp i\pi \rho(z) \qquad {\rm for} \quad z\in
[a,b].$$  

As we noted above, the two functions $\Om_g$ and $\Om_h$ obey linear
equations, obtained by differentiation of the equation \rf{elinear}\ satisfied by
$\om$. 
Following a method introduced in ref.~\cite{rEyZj}\ we can then easily determine them from their analyticity properties, and the boundary conditions.
$\Om_g(z)$ verifies the linear equation:
$$\Om_g(z+i0)+\Om_g(z-i0)=0$$
and behaves as $1/z$ when $z\to\infty$, and as $1/\sqrt{z-a}\sqrt{z-b}$ near
the cut end-points $a$ and $b$, because $\om$ behaves as
$\sqrt{z-a}\sqrt{z-b}$.  These conditions determine $\Om_g$ uniquely:
$$\Om_g(z,\l)={1\over \sqrt{(z-a)(z-b)}}.$$
 The same method applies to $\Om_h$ which satisfies
$$\Om_h(z+i0)+\Om_h(z-i0)=1/(\l-z),$$
and behaves like $\Om_h\sim O(1/z^2)$ for $z$ large, $\Om_h\sim
1/\sqrt{z-a}\sqrt{z-b}$ near $a,b$, and is regular near $z=\l$. It follows:
$$\Om_h(z,\l)={1\over2}{1\over \sqrt{(z-a)(z-b)}}\left( 1-{
\sqrt{(z-a)(z-b)}-\sqrt{(\l-a)(\l-b)}\over z-\l} \right).$$ 
Then if we set $z=\l$:
$$\Om_h(z=\l)={1\over 2\sqrt{(\l-a)(\l-b)}}-{1\over 2}{\d\over
\d\l}\ln\sqrt{(\l-a)(b-\l)}.$$
\medskip
We now have the necessary ingredients to determine $\partial F/ \partial
\l$: 
$$ -n^2{\partial F\over \partial \l}= N\om_0(z=\l) + (n-N) \Om_g(z=\l) +
\Om_h(z=\l) + O(1/N).$$ 

We still have to integrate all these terms with respect to $\l$.
In order to integrate $ \om_0= V'/2 -i\pi\rho$, we introduce a primitive of $\rho(\l)$:
$$\zeta(\lambda)=-\pi\int_a^\lambda\d\lambda'\rho(\lambda').$$ 

We also need a primitive of $\Om_g$.
For this purpose, we parametrize {\hbox{$\l=\oh (a+b)-\oh(b-a)\cos\varphi$}}, so that $\Om_g\propto 1/\sin\varphi$ and its primitive is simply
$$\int_a^\l \Om_g(\l') \d\l,=\varphi  \qquad {\rm where} \quad  \l=\ud(a+b)-\oh(b-a)\cos\varphi \qquad .$$
Finally, the result takes the form:
$$\begin{array}{rl}
-n^2 F(\l\pm i0) & =\oh NV(\l)\pm iN\zeta(\l) \mp i(N-n)\varphi \pm
i\ud\varphi\cr
 & \quad -\oh\ln{ \sqrt{(\l-a)(b-\l)}} + {\rm const}\ .\cr
\end{array}$$ 
Since $P_n$ is a polynomial, we have $P_n(\l)=\oh\left[ P_n(\l+i0)+P_n(\l-i0)
\right]$\ (this comes from the fact that there are actually two saddle points giving contributions of the same importance), and therefore: 
\beq\label{ePnasym}
P_n(\l)=\sqrt{2\over\pi}\, \ee{NV(\lambda)/2}{1\over\sqrt{f(\l)}}
\cos{\left[
N\zeta(\l)-(N-n)\varphi(\l)+\ud\varphi(\l)+{\rm const} \right] } ,
\eeq
where $f(\l)= \sqrt{(\l-a)(b-\l)}$.
The constant factor $\sqrt{2/\pi}$ is fixed by the condition that
$\kappa(\l,\l)=\rho(\l)$.
In the case of even potentials $V$, parity considerations imply that the
arbitrary constant phase is ${\rm const}=-\pi/4$.
 For general potentials the constant phase remains undetermined at this order, but we note that the general form of $P_n$ does not depend on the parity properties of~$V$.\par
We have thus proven the ansatz needed by \cite{rBrZe}\ .
\subsection{Connected Correlation Functions}

From this asymptotic expansion of $P_n$ one can now derive the kernel \rf{eKDC}\
$\kappa(\l,\mu)$, and then the connected correlation functions, in the
large $N$ limit. The authors of ref.~\cite{rBrZe}\ have calculated some correlation
functions in two regimes: short range correlations $(\l_i-\l_j)\sim 1/N$, and
mesoscopic correlations \hbox{$(\l_i-\l_j)\gg 1/N$.}  
Note that the polynomials $P_N$ and $P_{N-1}$ oscillate at a frequency of
order $N$ (which corresponds to the discrete spectrum of a matrix of size $N\
\times N$ finite), and therefore, all the correlation functions will present such
oscillations.   

In the short distance regime, these oscillations give the dominant behaviour,
and eq.~\rf{ePnasym},\rf{eKDC}\ lead to: 
$$\kappa(\l,\mu)\sim {\sin{[\pi N (\l-\mu)\rho(\l)]}\over \pi N(\l-\mu)}$$
and all connected correlation functions follow from \eq{erocnkappa}\ .
 All these features are studied in detail in
ref.~\cite{rBrZe}\ , and we will now concentrate our attention on the mesoscopic case.

In the regime $\l_i-\l_j\gg 1/N$, it is interesting to
consider smoothed functions, defined by averaging the fast oscillations.
For instance we find that:
\beq\label{erhodeu}
[ \roc_2(\l,\mu)]_{\rm smooth}={-1\over 2N^2\pi^2}{1\over
(\l-\mu)^2}{1-\cos\varphi \cos\psi\over \sin\varphi\sin\psi}\,,
\eeq  
where
\beq\label{epara}
\l={a+b\over2}-{b-a\over2}\cos\varphi \,,\quad
 \mu={a+b\over2}-{b-a\over2}\cos\psi\,. 
\eeq

Br\'ezin and Zee noted (ref.~\cite{rBrZe}) that the smoothed higher order $n$-points correlation functions vanished identically at the order $1/N^n$ for $n>2$ (Note that $\kappa$ being of order $1/N$, the connected $n$-point function will be of order $1/N^n$) . 
Indeed, we will prove below by a direct method, i.e. without smoothing \eq{erocnkappa}\ , that they are of order $1/N^{2(n-1)}$. The method is based on the study of the loop correlators, and these correlators have already been calculated recursively by the authors of \cite{rAJM}\ by the loop equation method. 
\subsection{The $n$-Loop Correlation Functions}
 
Let us consider the following functions:
\beq\label{edefloop}
 \om_n(z_1,\ldots ,z_n)=N^{n-2}\left< \tr {1\over
z_1-M}\times\ldots\times\tr {1\over z_n-M}\right>_{\rm conn}
\eeq 
They are related to the previous correlation functions by the relations: 
\beq\label{eomroc}
\om_n(z_1,\ldots ,z_n)=N^{2n-2}\int \prod_{i=1}^n {\d\l_i\over z_i-\l_i}
\,
\roc_n(\l_1,\ldots,\l_n)
\eeq
$\om_n$ is called the n-loop correlation function, because it is the
Laplace-transform of the partition function of a discrete random surface
limited by $n$ loops (the $z_i$ are conjugated to the lengths of the loops, see \cite{rBBIPZ,rDSBKGM,rAJM} ).
This remark allows to understand the topological origin of the factor $N^{2n-2}$: indeed,
the Laplace-transform of the complete $n$-point correlation function which is
of order $1$, would be the partition function of every surface (not necessary
connected) with $n$ boundaries.
Each surface contributes with a topological weight $N^\chi$ where $\chi$ is
its Euler character.
The leading term is the most disconnected one, with $\chi=n$ (indeed, such a surface is made of $n$ discs, each of them having $\chi=1$), while the connected term has $\chi=2-n$ (it is a sphere ($\chi=2$) from which $n$ discs have been removed).
Therefore, the relative contribution of the connected part to the complete $n$-loop function is $N^{2-2n}$.
As stated before, one can not only find the large $N$ scaling of the connected
correlation functions, but also compute them exactly by the recursive method
of \cite{rAJM}.
Let us give a brief summary of this method in the case which we are interested in.

First remark that relation \rf{eomroc}\ can be inverted: $\om_n$ is analytical except when some of the $z_i$ belong to the interval $[a,b]$.
$\roc_n$ can then be expressed in terms of the differences of $\om_n$ between opposite sides of the cut.
For instance:
$$\begin{array}{rl}
 \rho(\l) & ={-1\over 2i\pi}(\om(\l+i0)-\om(\l-i0))\cr
\roc_2(\l,\mu) & ={1\over
(2i\pi N)^2}(\om_2(\l+i0,\mu+i0)-\om_2(\l+i0,\mu-i0)-\om_2(\l-i0,\mu+i0) \cr
 & \quad +\om_2(\l-i0,\mu-i0)).\cr
\end{array}$$
and for general $n$:
\beq\label{erocom}
\roc_n(\l_1, \dots,\l_n)={1\over N^{2n-2}}\left({-1\over 2i\pi}\right)^n
\sum_{\epsilon_i=\pm 1}
\left(-1\right)^{(\epsilon_1+\dots+\epsilon_n)}\, \om_n(\l_1+\epsilon_1
i0,\ldots,\l_n+\epsilon_n i0)
\eeq
Note that these functions are directly the smoothed correlation functions,
since we first compute $\om_n$ in the large $N$ limit at complex arguments
(which suppresses the oscillations), and then take the discontinuities along
the cut.  

The main tool of \cite{rAJM}\ is the loop-insertion-operator, which acts on the
free-energy of the Matrix-model, and gives the loop-correlation-functions:
consider the partition function:
$$ Z= \e^{N^2 F} = \int \d M\, \ee{-N\tr  V(M)} $$
with $V(z)=\sum_{k=1}^\infty g_k z^k/k$, and define the loop-insertion-operator: 
\beq\label{eloopinsert}
 {\delta\over \delta V(z)}=-\sum_{k=1}^\infty {k\over
z^{k+1}} {\partial\over\partial g_k}.
\eeq
Note that with this definition
\beq\label{eVder}
{\delta V(z')\over \delta V(z)}={1\over z-z'}-{1\over z}\,.
\eeq
The $\om_n$ are obtained from the free energy $F$ by the repeated action of
this operator: 
\bea\label{eloopn}
\om_n(z_1, \ldots, z_n) = {\delta\over\delta V(z_1)} \cdots {\delta\over \delta V(z_n)}F \,, & \label{eloopna} \cr
 = {\delta\over\delta V(z_n)}\om_{n-1}(z_1, \ldots, z_{n-1}). & \label{eloopnb}\cr
\eea
It is not necessary to calculate the free energy $F$, since we already  know
the one-loop function $\om(z)$.  
We have already emphasized that $\om(z)$ satisfies a linear equation
\eq{elinear}, and thus, all its derivatives satisfy the same linear
equation, with a different  l.h.s. .
Then, analyticity properties, and boundary conditions determine the form of $\om_n$. 

The linear equation for $\om$ is:
\beq\label{elooplin}
 \om(\l+i0)+\om(\l-i0)=V'(\l).
\eeq
By a repeated action of the loop insertion operator, we obtain:
\beq\label{eloopdeux}
 \om_2(\l+i0,z)+\om_2(\l-i0,z)=-{1\over (z-\l)^2},
\eeq
and for $n>2$:
\beq\label{enloop}
 \om_n(\l+i0,z_2,\ldots,z_n)+\om_n(\l-i0,z_2,\ldots,z_n)=0\,.
\eeq
The function $\om(z)$ has the form
$$ \om(z)=\ud\left( V'(z)-M(z)\sqrt{(z-a)(z-b)}\right)$$
where $M(z)$ is a polynomial such that $\om(z)\sim 1/z$ when $z\to\infty$,
and therefore:
$$M(z)=\left({V'(z)\over \sqrt{(z-a)(z-b)}}\right)_+ \quad .$$
Then
$$\rho(\l)={1\over 2\pi}M(\l)\sqrt{(\l-a)(b-\l)}={1\over
2\pi}M(\l){b-a\over2}\sin\varphi\,,$$ 
where we have used the parametrization \rf{epara}\ .\par
The two-loop function is also completely determined by the linear equation
\rf{eloopdeux}\ and boundary conditions: 
$$ \om_2(x,y)=-{1\over 4}{1\over (x-y)^2}\left(
2+{(x-y)^2-(x-a)(x-b)-(y-a)(y-b)\over
\sqrt{(x-a)(x-b)}\sqrt{(y-a)(y-b)}}\right)$$  
and thus, in agreement with \eq{erhodeu}:
$$ \roc_2(\l,\mu)=-{1\over 2N^2\pi^2}{1\over
(\l-\mu)^2}{1-\cos\varphi\cos\psi\over \sin\varphi\sin\psi}.$$ 

The other loop functions all satisfy an homogeneous equation, and can be
written: 
$$ \om_n(\l_1,\ldots,\l_n)=\left( \prod_{i=1}^n {1\over
\sin{\varphi_n}}\right)^{2n-3} W_n(\l_1,\ldots,\l_n),$$
with now $\l_n={a+b\over2}-{b-a\over2}\cos\varphi_n$, and
where the $W_n$ are some symmetric polynomials of degree less than $2n-5$ in
each $\l_i$, which are no longer determined by the boundary conditions. 
It is necessary to directly use the recursion relation \rf{eloopnb}.
Since $\omega_2$ depends on the potential $V(M)$ only through $a$ and $b$, 
we need the actions of loop-insertion operator on $a$ and $b$, for instance 
$ {\delta a/\delta V(z)}$,~${\delta^2 a/\delta V(z)\,\delta V(z')}$.... For
this purpose, the authors of \cite{rAJM}\ introduced the following moments of the potential: 
$$ M_k=-{1\over 2i\pi} \oint\d z\, {1\over (z-a)^k} {V'(z)\over
\sqrt{(z-a)(z-b)}}$$ 
$$ J_k=-{1\over 2i\pi} \oint\d z\, {1\over (z-b)^k} {V'(z)\over
\sqrt{(z-a)(z-b)}}$$
(the integration path turns clockwise around the cut $[a,b]$).

They are such that:
$$M(z)=\sum_k M_{k+1} (z-a)^k=\sum_k J_{k+1} (z-b)^k$$
The $M_k$ and $J_k$ are linearly related to the coefficients of the
potential $V$, and if $V$ is a polynomial of degree $v$, $M$ is of degree
$v-2$, and there are only $v-1$ independent coefficient among the $M_k$ and
$J_k$. Note also, that if $V$ is even, we have $M_k=-(-1)^k J_k$.  

The cut end-points $a$ and $b$ depend on $V$ through the conditions that (see
\cite{rAJM,rDGZ}):
$$ \begin{array}{rl}
 {1\over 2i\pi}\oint\d z\, {V'(z)\over \sqrt{(z-a)(z-b)}}
 & =0\,,\cr 
{1\over 2i\pi}\oint\d z\, {zV'(z)\over \sqrt{(z-a)(z-b)}} & =2\,.\cr 
\end{array}$$ 
It follows (using \eq{eVder} and performing the contour integrals):
$$\begin{array}{rl}
{\delta a\over\delta V(z)} & = {1\over M_1}{1\over z-a}{1\over
\sqrt{(z-a)(z-b)}}\,, \cr 
{\delta b\over\delta V(z)} & = {1\over J_1}{1\over z-b}{1\over
\sqrt{(z-a)(z-b)}}\,.\cr
\end{array}$$ 
In order to determine the higher order derivatives of $a$ and $b$, we need to
differentiate the coefficients $M_k$ and $J_k$: 
$$\begin{array}{rl}
{\delta M_k\over \delta V(z)} = & {2k+1\over2} {\delta a\over\delta
V(z)}\left(M_{k+1}-{M_1\over (z-a)^k}\right) \cr
 & + {1\over2}{\delta b\over \delta V(z)}\left({J_1\over (b-a)^k}-{J_1\over
(z-a)^k}-\sum_{l=0}^{k-1}{M_{l+1}\over (b-a)^{k-l}} \right) \cr 
\end{array}
$$
and analogous formulae are obtained for the $J_k$ by the exchange $a\leftrightarrow b$.

With these tools, we can now determine the $\om_n$ recursively. 
For instance $\om_3$ is given in \cite{rAJM}\ :
$$\begin{array}{rl}
\om_3(x,y,z) = & {a-b\over 8\left(\s{x}\s{y}\s{z}\right)^3}               \cr
               & \quad\times\left( {1\over M_1} (x-b)(y-b)(z-b) - {1\over J_1} (x-a)(y-a)(z-a) \right)                                                   \cr
\end{array}$$
Let us also write $\omega_4$. For this purpose, we first define the polynomials
$$Q(x_i,a)=\prod_{i=1,4}\left(x_i-a\right).$$
With this notation
$$\begin{array}{rl}
\om_4(x_i) = & {-1 \over16 \left[Q(x_i,a)Q(x_i,b)\right]^{3/2}}
\left( -3(b-a){M_2\over M_1^3} Q(x_i,b)-3(a-b){J_2\over J_1^3}Q(x_i,a)
\right.                                                                   \cr 
             & \quad +3{1\over M_1^2}Q(x_i,b)\left[
(b-a)\left(\sum\nolimits_{i=1}^4{1\over x_i-a}\right) -1\right]           \cr 
             & \quad +3{1\over J_1^2}Q(x_i,a)\left[
(a-b)\left(\sum\nolimits_{i=1}^4{1\over x_i-b}\right) -1\right]           \cr 
             & \left.\quad+{1\over M_1 J_1} \left[(x_1-a)(x_2-a)(x_3-b)(x_4-b)+\ {\rm 5\ terms}  \right] \right),                                          \cr
\end{array} $$ 
where the last additional terms symmetrize in the four variables.
%& -16\om_4(x,y,z,t)=\cr
%& -3(b-a)\left( M_2 M_1 a_x a_y a_z a_t- J_2 J_1 b_x b_y b_z b_t \right) \cr
%&+3 M_1^2 a_x a_y a_z a_t \left[ (b-a)\left( {1\over x-a}+{1\over
%y-a}+{1\over z-a}+{1\over t-a}\right)-1\right]\cr 
%&+3 J_1^2 b_x b_y b_z b_t \left[ (a-b)\left( {1\over x-b}+{1\over
%y-b}+{1\over z-b}+{1\over t-b}\right)-1\right]\cr 
%&+ M_1 J_1 \left[ a_x a_y b_z b_t+ a_x b_y a_z b_t+ b_x a_y a_z b_t\right.
%&+ \left. a_x b_y b_z a_t+ b_x a_y b_z a_t+ b_x b_y a_z a_t\right] \cr } $$ 
The connected functions $\roc_n$, are then obtained by \eq{erocom}\ and \eq{enloop},
and they are simply given by:
$$\roc_n(\l_1,\dots,\l_n)={1\over N^{2n-2}}\left( {-1\over i\pi}\right)^n
\om_n(\l_1+i0,\dots,\l_n+i0).$$

\subsection{Universality of the smoothed correlation functions}

- We observe that the only universal features of ${\cal O}_1(\lambda)=\rho(\lambda)$ are the square-root singularities at the edge of
the distribution, otherwise the function $\rho(\l)$ is potential dependent.

- The two point function is universal.

- The $n$-point smoothed correlation functions with $n\geq 3$ can be calculated
recursively by a systematic method \cite{rAJM}.
 The main property is that they consist in a sum of a finite number of universal functions and involve only the first $n-2$ moments of the potential.
That means for instance, that two potentials $V$ and $V^*$ induce the same three-point function as soon as they yield  the same $M_1$ and $J_1$ coefficients, but they don't need to be identical.
However, the $n$-point correlation function is no longer given by a gaussian model, since it is of order $1/N^{2n-2}$.
Perturbative corrections to the gaussian model have to be considered.
\smallskip
Note that the determination of correlation functions allows an evaluation
of the moments $M_k$, in a case where the potential $V$ is unknown.
Let us recall \cite{rAJM,rACKM} that these moments play an important role in the study of critical points.
Since
$$\rho(\l)={1\over 2\pi}M(\l)\sqrt{(\l-a)(b-\l)}$$
we see that if some moments vanish, the behaviour of $\rho$ near the
end-points is no longer a square-root.
When the $m$ first $M_k$ vanish, one finds $\rho\sim (\l-a)^{m+1/2}$ which corresponds to the $m^{\rm th}$ multicritical point of the one-matrix model of 2D gravity \cite{rDSBKGM,rDGZ}.

\subsection{Conclusions}

In this section we have recovered, by a completely different method, the
results of ref.~\cite{rBrZe}\ concerning the two-point eigenvalue correlation
function of a random hermitian matrix in the limit in which the size $N$ of
the matrix becomes large.
This method of orthogonal polynomials will be extended to another model in the next section.
However, this method is not convenient to evaluate smoothed correlation functions.
The way to do that, is the recursive method of \cite{rAJM}.
Note that this method can easily be extended to the $O(n)$ matrix model using the two-point function and the genus expansion procedure presented in \cite{rEyZj,rEyKr}.

Br\'ezin and Zee in ref.~\cite{rBrZe}\ have shown that the two-point function is
universal, and therefore identical to the function of the gaussian matrix
model.
The gaussian model is the fixed point of a renormalization group
\cite{rBrZi,rHINS} and a direct RG analysis should be performed to put this
result in perspective. 

In the same way, Br\'ezin and Zee have shown that higher smoothed correlation functions vanish at leading order.
The contributions calculated by the loop insertion method of \cite{rAJM}\ should be considered as corrections to the leading scaling behaviour.
The explicit expressions show that they depend now on successive moments of the potential, indicating an implicit classification of the deviations from the gaussian model in terms of their irrelevance for $N$ large.
Here also it would be interesting to confirm the qualitative aspects of these
results by a direct RG analysis.

\newsection{ The 2-matrix model}

The previous method to evaluate the correlation functions of the eigenvalues of a single random matrix can be extended to the case of a multi-matrix model.
It is known from 2D quantum gravity \cite{rDGZ}\ that the multi-matrix model provides a larger class of critical points, and therefore, the universality properties may be different from the one-matrix case.
It has also been established \cite{rDKK}\ that a two-matrix model is sufficient to reproduce all the critical behaviours of multi-matrix model, we shall therefore concentrate our attention on this case, but all the method may be generalized to any chain of matrices, provided you can use the Itzykson-Zuber integration to define orthogonal polynomials \cite{rCorreln}.

A multi-matrix model can be regarded as a particular one-matrix model with a non-polynomial potential, if you integrate over all the other matrices.
In this case, you are interested in the correlation functions of the eigenvalues of only one of the matrices, and we will see below that in the large $N$ limit, we recover the same universal behaviour as for one matrix with polynomial potential.
The multi-matrix model presents also some interest in the study of coupled systems \cite{rBreZee,rAlt}.
In that case, you are interested in the correlations of eigenvalues of different matrices.
We will see, that in the large $N$ limit, the eigenvalues of the two matrices are essentially uncorrelated.

\medskip

First, let us define the model and the problem.
We will try to use the same presentation and notations as for one matrix:\par
we consider two stochastic matrices $M$ and $\t{M}$ hermitian, of size $N\times N$, with a probability:
\beq\label{eprobdeu}
{\cal P}(M,\t{M})={1\over Z} \ee{-N\tr (V(M)+\t{V}(\t{M})-M\t{M})}
\eeq
Where $V$ and $\t{V}$ are two given polynomial potentials, not necessarily of the same degree.
This probability distribution induces by integration over the angular degrees of freedom of the matrices (the unitary group) a joint probability density for the eigenvalues of both matrices.
Using the famous formula of \cite{rIZ}\ , the result reads:
\beq\label{eprobvaldeu}
{\cal P}(\l_i,\t\l_j)={1\over Z} \Delta(\l)\Delta(\t\l) \det (\ee{N\l_i\t\l_j}) \prod_i \ee{-N(V(\l_i)+\t{V} (\t\l_i))}
\eeq
where the $\Delta$ are the Vandermonde determinants:
$$\Delta(\l)=\prod_{i<j}(\l_i-\l_j)\qquad , \qquad \Delta(\t\l)=\prod_{i<j}(\t\l_i-\t\l_j)$$
The quantities we are interrested in are the correlation functions of eigenvalues.
We consider the operators:
\beq\label{edefO}
 \begin{array}{rll}
 \O(\l)     & = {1\over N}\tr \delta (\l-M) & = {1\over N}\sum_i \delta(\l-\l_i)      \cr
 \t\O(\t\l) & = {1\over N}\tr\delta(\t\l-\t{M}) & = {1\over N}\sum_i \delta(\t\l-\t\l_i) \cr
\end{array}
\eeq
One would like to evaluate mean values as:
$$ \O_1(\l)=\moy{\O(\l)}$$
$$ \t\O_1(\t\l)=\moy{\t\O(\t\l)}$$
$$ \O_{2,0}(\l,\mu)=\moy{\O(\l)\O(\mu)}_{\rm conn}$$
\beq\label{edefOnm}
 \O_{n,m}(\l_1,\dots,\l_n,\t\l_1,\dots,\t\l_m)=\moy{\prod_{i=1}^n \O(\l_i) \,\prod_{i=1}^m \t\O(\t\l_i)}_{\rm conn} \,  .
\eeq
These correlations are related to the partially integrated eigenvalue distributions:
\beq\label{eronmpid}
 \rho_{n,m}(\l_1,\dots,\l_n,\t\l_1,\dots,\t\l_m)=\int\dots\int \prod_{i>n} \d{\l_i} \prod_{j>m} \d{\t\l_j} {\cal P}(\l_1,\dots,\l_N,\t\l_1,\dots,\t\l_N)
\eeq
Indeed, when all the $\l_1,\dots,\l_n$ and the $\t\l_1,\dots,\t\l_m$ are distincts, we have:
$$ \O_{n,m}={N!\over N^n(N-n)!}{N!\over N^m(N-m)!}\rho^{(c)}_{n,m}$$

Unlike the one-matrix case, we have here three kinds of correlation functions: those with only eigenvalues of $M$, those with only $\t{M}$, and the mixed ones.
For instance, the densities of eigenvalues:
$$\rho(\l) \qquad {\rm and} \qquad \t\rho(\t\l) \quad ,$$
 and the two point-functions:
$$\rho_{2,0}(\l,\mu) \quad \rho_{0,2}(\t\l,\t\mu) \quad \rho_{1,1}(\l,\t\l)$$
From the one-matrix model case, we guess that the function $\rho_{2,0}(\l,\mu)$ shows an interesting behaviour at small distance $\l-\mu\sim 1/N$, but what about the mixed correlation ? $\l$ and $\t\l$ are not quantities of the same nature ($\t\l$ scales as $1/\l$), and nothing will happen when $\l$ is close to $\t\l$ (except perhaps in the symetric case $V=\t{V}$).
Also, as in the one-matrix case, another interresting problem is to calculate the smoothed correlations at large distances ( $\gg 1/N$).

As in the one-matrix case, the method of orthogonal polynomials provides the short distance correlations in terms of some kernels (here there are four kernels), and these expressions can be smoothed to obtain the smoothed two-point functions.
The higher order smoothed correlation functions can't be obtained by this method, but by the loop equations, which are too complicated in the two-matrix-model, i.e. there is not yet any generalization of \cite{rAJM}.

First, we recall the method of orthogonal polynomials and introduce the kernels.

\subsection{orthogonal polynomials}

We will consider two families of polynomials $P_n$ and $\t{P}_n$ ($n$ is the degree) with the same leading term, and which satisfy:
\beq\label{eorthodeu}
 \int \d\l \d{\t\l} P_n(\l)\t{P}_m(\t\l) \ee{-N(V(\l)+\t{V}(\t\l)-\l\t\l)} =\delta_{n,m}
\eeq
We also define the wave functions:
\beq\label{edefpsi}
\begin{array}{rl}
 \psi_n(\l)     & = P_n(\l)\ee{-NV(\l)}                 \cr
 \t\psi_n(\t\l) & = \t{P}_n(\t\l)\ee{-N\t{V}(\t\l)}     \cr
\end{array}
\eeq
Which we note:
\beq\label{edefbra}
\begin{array}{rl}
 <n|    & = \psi_n(\l)<\l|            \cr
|\t{n}> & = \t\psi(\t\l)|\t\l>        \cr
\end{array}
\eeq
they form two dual spaces with the bilinear form:
$$<n|\t{m}>=\int \d\l \d{\t\l} \psi_n(\l) \t\psi_m(\t\l) \ee{N\l\t\l}=\delta_{n,m}$$
with this duality, it is possible to map each of these spaces into the other one, we thus define two other families of functions (which are not polynomials):
\beq\label{edefchi}
\begin{array}{rl}
 \chi_n(\l)     & = \int \d{\t\l} \t\psi_n(\t\l) \ee{N\l\t\l}     \cr
 \t\chi_n(\t\l) & = \int \d\l \psi_n(\l) \ee{N\l\t\l}             \cr
\end{array}
\eeq
As in the one matrix-case, partial integrations over eigenvalues in \eq{eprobvaldeu}\
are performed by expressing the Vandermonde determinants in terms of orthogonal polynomials, and result into combinations of four kernels:
\beq\label{edefKer}
\begin{array}{cc}
 H(\l,\mu) = {1\over N}\sum_{k=0}^{N-1} \psi_k(\l)\chi_k(\mu)
\quad & \quad
K(\l,\t\l) = {1\over N}\sum_{k=0}^{N-1} \psi_k(\l)\t\psi_k(\t\l)\ee{N\l\t\l} \cr
 \t{H}(\t\l,\t\mu) = {1\over N}\sum_{k=0}^{N-1} \t\chi_k(\t\l)\t\psi_k(\t\mu)
\quad & \quad
 \t{K}(\t\l,\l) = {1\over N}\sum_{k=0}^{N-1} \t\chi_k(\t\l)\chi_k(\l)\ee{-N\t\l\l} \cr
\end{array}
\eeq
We then have (see Appendix 4 or \cite{rMehtaker}):
\beq\label{erhoKer}
 \rho(\l)=H(\l,\l)\qquad
 \t\rho(\t\l)=\t{H}(\t\l,\t\l)
\eeq
\beq\label{erhodeuKer}
\begin{array}{rl}
 \rho^{(c)}_{2,0}(\l,\mu)  & = -H(\l,\mu)H(\mu,\l)   \cr
 \rho^{(c)}_{1,1}(\l,\t\l) & = -K(\l,\t\l)\t{K}(\t\l,\l)+{1\over N}K(\l,\t\l)                \cr
\end{array}
\eeq
As for one matrix, we have to evaluate these kernels, this can be done by a generalisation of the Darboux-Christoffel theorem.
We have for instance:
\beq\label{eDCderK}
\left(\l+{1\over N}{\partial\over \partial\t\l}\right) K(\l,\t\l)\ee{-N\l\t\l}={1\over N} \left(
\alpha \psi_N\t\psi_N-\sum_{k=1}^{{\rm deg}\t{V}'}\sum_{i=0}^{k-1} \psi_{N-k+i}\t\psi_{N+i}\right)
\eeq
wich involves only $\psi_n$ and $\t\psi_n$ with $n-N\sim 1$.
Note that this is a differential equation for $K$, but in the large $N$ limit, the partial derivative will be replaced by a multiplication by a function $\l(\t\l)$, as we will see below.
Such Darboux-Christoffel like formulaes exist also for the other kernels.
Therefore, our next task will be to find an asymptotic expression for the $\psi_n$ in the limit $N$ large and $n-N\sim 1$.

Let us now explain the method (similar to the one \cite{rDKK}\ used to calculate the two-loop functions).
We consider the following operators of our Hilbert spaces:
\beq\label{eop}
 \hat\l \quad , \quad\hat{P}={1\over N}{d\over d\l}\qquad;\qquad \hat{\t\l}\quad ,\quad \hat{\t{P}}={1\over N}{d\over d\t \l}\qquad ,
\eeq
$$ \hat{n} \quad , \hat{\phi}={d\over dn} \quad ,\quad \hat{x}=\ee{-\hat\phi} $$
such that:
$$<n|\hat\l = \l\psi_n(\l)<\l|$$
$$\hat{\t\l}\ket{\t{n}}= \t\l \t\psi_n(\t\l) |\t\l>$$
$$\hat{n}|\t{m}>=m|\t{m}> \quad , \quad \hat{x}|\t{n}>=|\t{n-1}>$$
$$<m|\hat{n}=m<m| \quad ,\quad  <n|\hat{x}=<n+1|$$
It is easy to see from the degrees of the polynomials that $\hat\l$ and $\hat{\t\l}$ have the form (for instance, a multiplication by $\l$ can raise the degree of at most 1):
\beq\label{elalphakop}
\begin{array}{rl}
 \hat\l     & = \alpha(\hat{n})\hat{x}+\sum_{k\geq 0} \alpha_k(\hat{n})\hat{x}^{-k}                               \cr
 \hat{\t\l} & = \hat{x}^{-1}\alpha(\hat{n})+\sum_{k\geq 0} \hat{x}^k \t{\alpha}_k(\hat{n})                                       \cr
\end{array}
\eeq
the equations of motion ( just integrate \eq{eorthodeu}\ by parts) give simply:
\beq\label{eqPl}
\hat{P}=-\hat{\t\l} \qquad , \quad \hat{\t{P}}=-\hat{\l}
\eeq
 and:
\beq\label{eqmotion}
\begin{array}{rl}
 \hat{P}     & = -V'(\hat\l) + {\hat{n}\over N}{1\over \hat{x}}{1\over \alpha(\hat{n})} + O({1\over \hat{x}^2})                                 \cr
 \hat{\t{P}} & = -\t{V}'(\hat{\t\l}) + {1\over \alpha(\hat{n})} \hat{x} {\hat{n}\over N} + O(\hat{x}^2)                                          \cr
\end{array}
\eeq
The expansion of \eq{eqmotion}\ in power series of $\hat{x}$, using commutation relations like 
$$[\hat{x},\alpha(\hat{n})]=(\alpha(\hat{n}+1)-\alpha(\hat{n}))\hat{x}$$
gives a set of recursion relations between the $\alpha_k$,$\t\alpha_k$.
these last equations allow (in principle) to determine all the $\alpha_k$ and $\t\alpha_k$, by induction. Note that only a finite number of $\alpha_k$ are non zero, and more precisely:
$$\hat\l=\alpha(\hat{n})\hat{x}+\sum_{k= 0}^{{\rm deg} \t{V}'} \alpha_k(\hat{n})\hat{x}^{-k}$$
$$\hat{\t\l}=\hat{x}^{-1}\alpha(\hat{n})+\sum_{k= 0}^{{\rm deg} V'} \hat{x}^k \t{\alpha}_k(\hat{n})$$
Another remark is that all these recursion equations are equivalent to the cannonical commutation relations:
\beq\label{ecommut}
 [\hat\l , \hat{\t\l}]={1\over N}
\eeq
Up to now we have written only exact equations, let us now discuss the large $N$ limit approximation.
\medskip

\subsection{Large N limit}

It is the classical limit, i.e. we can drop the commutators, and replace
operators by numbers.
We will also be interested in the limit $n-N\sim 1$, and we will most of time replace the operator $\hat{n}$ by the value $N$.
The $\alpha_k(\hat{n})$ are then numbers $\alpha_k$ which no more depend on $n$.
We thus write:
\beq\label{elclassic}
\begin{array}{rl}
 \l    & = \alpha x + \sum_{k= 0}^{{\rm deg} \t{V}'} \alpha_k x^{-k}    \cr
 \t\l  & = \alpha x^{-1} + \sum_{k= 0}^{{\rm deg} V'} \t\alpha_k x^k    \cr
\end{array}
\eeq
At leading order the recursion equations for the $\alpha_k$ become algebraic equations, they come from the expansion in power of $x$ of the classical equations of motion:
\beq\label{eqmotioncl}
\begin{array}{rl}
 V'(\l) - \t\l     & = {1\over x}{1\over \alpha} + O({1\over x^2})     \cr
 \t{V}'(\t\l)-\l   & = x {1\over \alpha} + O(x^2)                      \cr
\end{array}
\eeq
still, we remark that these algebraic equations can also be obtained from a Poisson bracket:
\beq\label{epoisson}
 x\left({\partial \l\over \partial n}{\partial \t\l \over \partial x}-{\partial \l\over \partial x}{\partial \t\l \over \partial n}\right)=\left\{ \l , \t\l \right\} = {1\over N}
\eeq
The left member is expanded in powers of $x$. In particular the coefficient of $x^0$ is:
$${d\over dn}\left(\alpha^2-\sum_k k\alpha_k\t\alpha_k\right)=-{1\over N}$$
which gives after integration:
\beq\label{ealphasq}
 \alpha^2-\sum_k k\alpha_k\t\alpha_k=-{n\over N}=-1
\eeq
From these algebraic equations, one determines all the coefficients $\alpha_k$, and obtains two functions $\l(x)$ and $\t\l(x)$ of an auxillary variable $x$:
\beq\label{elx}
\begin{array}{rl}
\l(x)   & = \alpha x + \sum_{k\geq 0} \alpha_k x^{-k}       \cr
\t\l(x) & = \alpha x^{-1} + \sum_{k\geq 0}\t\alpha_k x^k    \cr
\end{array}
\eeq
In particular, we can define a function $\t\l(\l)$: we fisrt get $x$ from the first relation and insert it into the second one.
Such a function is multivaluate, it has a cut along a segment $[a,b]$ of the complex plane.
Actually, it can be seen from the loop equations (see appendix 6 or \cite{rSt} ) that $V'(\l)-\t\l(\l)=\om(\l)$ is precisely the resolvent:
$$ \om(z)={1\over N}\moy{ \tr {1\over z-M} }$$

\smallskip
We now arrive to the aim of those calculations, i.e. finding an asymptotic approximation of the orthogonal polynomials, and kernels.

The leading behaviour of $\psi_n(\l)$ when $N\gg 1$ (classical limit) and $n\sim N$, is simply given by:
\beq\label{edifpsin}
 {1\over N}{d\psi_n(\l)\over d\l}\sim -\t\l(\l) \psi_n(\l)
\eeq
indeed $\hat{P}=-\t\l$ according to \eq{eqPl}\ ,
and therefore:
$$ {d\ln{\psi_n(\l)} \over d\l} \sim -N\t\l(\l)$$
As in the one matrix case, we define:
\beq\label{edefzeta}
 \zeta(\l)=i\int \t\l(\l)\, \d\l
\eeq
and obtain:
$$ \psi_n(\l)\sim \ee{iN\zeta(\l)}$$
This rough approximation is not sufficient to determine the kernels, we have to go to order $O(N^0)$.
Remark that the dependance in $N-n$ is easily found, since:
$$<n|=<N|\hat{x}^{n-N}$$
or in other words:
$$ \psi_n=x^{n-N}\psi_N$$
A proof is presented in Appendix 5, we give here only the final result, which is very close to the one obtained for one matrix (remember \eq{ePnasym} ):
\beq\label{epsinapp}
\begin{array}{rl}
   \psi_n(\l)     & \sim \sqrt{2\over\pi} \sum_{x}\, {1\over \sqrt{\d\l\over \d{x}}} \ee{-N\int_{x_0}^x \t\l(y)\l'(y)\d{y}} x^{n-N}         \cr
   \t\psi_n(\t\l) & \sim \sqrt{2\over\pi} \sum_{x}\, {1\over \sqrt{-{\d{\t\l}\over \d{x}}}} \ee{-N\int_{x_0}^x \l(y)\t\l'(y)\d{y}} x^{N-n-1}
                                                               \cr
   \chi_n(\l)     & \sim \sqrt{2\over\pi} \sum_{x}\, {1\over \sqrt{-{\d\l\over \d{x}}}} \ee{N\int_{x_0}^x \t\l(y)\l'(y)\d{y}} x^{N-n-1}       \cr
   \t\chi_n(\t\l) & \sim \sqrt{2\over\pi} \sum_{x}\, {1\over \sqrt{\d{\t\l}\over \d{x}}} \ee{N\int_{x_0}^x \l(y)\t\l'(y)\d{y}} x^{n-N}          \cr
\end{array}
\eeq
{\it Remarks:}
- The $\sum_{x}$ means that you have to sum over all the values of $x$ which give the same value of $\l=\l(x)$ (in the one-matrix case, we had only two values of $x$ complex conjugate of each other ($x=\ee{\pm i\varphi}$), that's why the sum over $x$ was replaced by a real part).\par
- $\t\psi$ can be obtained from $\psi$ by the exchange $x\leftrightarrow 1/x$, and $\alpha_k\leftrightarrow\t\alpha_k$.\par
- The expressions of $\chi$ and $\t\chi$ are obtained by a steepest descent approximation in \eq{edefchi} .\par
- The lower bound of integration $x_0$ can be chosen arbitrary but must be the same for all.
It can be chosen for instance such that $\l'(x_0)=0$, and such that $\l(x_0)=a$ is one of the cut end-points of the cut $[a,b]$ of the resolvent $\om(z)$.

\subsection{Darboux-Christoffel theorem for the kernels}

We have already asserted, that there exists a kind of Darboux-Christoffel theorem, which gives the kernel \rf{edefKer} in terms of a litle number of $\psi_n$. For $K$ and $\t{K}$, we have only differential equations, but in the classical limit, the derivatives disappear.
For instance, we have:
$$ \begin{array}{rl}
 K(\l,\t\l) \ee{-N\l\t\l} & = {1\over N} \sum_{n=0}^{N-1}  \ket{\t{n}} \otimes \bra{n}                            \cr
                          & = {1\over N} \sum_{n=1}^\infty \hat{y}^n \ket{\t{N}} \otimes \bra{N} \hat{x}^{-n}       \cr
\end{array}$$
with the operators $\hat{x}$ and $\hat{y}$ acting on the $\bra{N}$ and $\ket{\t{N}}$ respectively, they commute, and we can sum up the geometrical serie:
\beq\label{eDCxyK}
 K(\l,\t\l)\ee{-N\l\t\l}={1\over N}{1\over {\hat{x}\over \hat{y}}-1}\ket{\t{N}}\bra{N}
\eeq
in the classical limit, $\hat{x}$ and $\hat{y}$ become $x=x(\l)$ and $y=x(\t\l)$.
Let us multiply this expression by $\l-\l(\t\l)$ which is:
$$ \l-\l(\t\l)=\l(x)-\l(y)=(x/y-1)\left(\alpha y-\sum_k \alpha_k \sum_{i=0}^{k-1} x^{i-k}y^{-i}\right)$$
Therefore:
\beq\label{eDClK}
 K(\l,\t\l)\ee{-N\l\t\l}={1\over N}{1\over \l-\l(\t\l)}\left(\alpha y-\sum_k \alpha_k \sum_{i=0}^{k-1} x^{i-k}y^{-i}\right) \ket{\t{N}}\bra{N}
\eeq
which is formally identical to \eq{eDCderK}.
By Laplace transforms, we can express the other kernels:
\beq\label{eDClH}
 H(\l,\mu)={1\over N}{1\over \l-\mu}\left( \alpha \psi_N(\l)\chi_{N-1}(\mu)-\sum_k \alpha_k \sum_{i=0}^{k-1} \psi_{N-k+i}(\l)\chi_{N+i}(\mu)\right)
\eeq
and analogous expressions for $\t{H}$ and $\t{K}$.

We can now insert \rf{epsinapp} in \rf{eDClK}... \rf{eDClH}, and compute the correlation functions \rf{erhoKer}, \rf{erhodeuKer}.

\subsection{Correlation functions}

Let us first compute the kernel $H(\l,\mu)$, and in particular the limit $\l=\mu$ which will allow to fix the normalizations.
From \rf{eDCxyK} we have:
$$ H(\l(x),\mu(y))={1\over N}{1\over x/y-1}\psi_N(\l)\chi_N(\mu)$$
and with \rf{epsinapp} 
$$ H(\l,\mu)={\rm cte}^2 \sum_{x,y} {1\over N}{1\over \sqrt{\d\l\over \d{x}}\sqrt{-{\d{\l}\over \d{y}}}} {1\over x-y}\ee{ N\int_x^y \t\l(\xi)\l'(\xi) \d\xi }$$
All the $x$ and $y$ such that $\l=\l(x)$ and $\mu=\l(y)$ contribute to the result in the general case.
But in the case $\l-\mu$ small, the terms such that $x-y$ is small are dominant , and there are two such terms, complex conjugate of each other.
We thus have:
$$ H(\l,\mu)\mathop\sim_{\l\to\mu} \Re {i\over N \pi}\,{1\over (\l-\mu)}\,\ee{-N\t\l(\l) (\l-\mu)}$$
Since we have $\om(\l)=V'(\l)-\t\l(\l)$, 
$$ -\Im \t\l(\l)=\Im \om(\l)=-\pi \rho(\l)$$
we recover $H(\l,\l)=\rho(\l)$.
(Actually, this relation was used to fix the prefactor $\rm cte=1/\sqrt{2\pi}$ in \rf{epsinapp} ).

The short distance ($\l-\mu$ small) expression of $H$ is identical to the one obtained in the 1-matrix  case:
$$ H(\l,\mu)\sim  {\sin{\pi N (\l-\mu)\rho(\l)}\over\pi N(\l-\mu)}\, \ee{-N(\l-\mu)\Re \t\l(\l)}$$
The two point function can then be derived:
\beq\label{erhodeumicro}
 \rho^{(c)}_{2,0}(\l,\mu) = -H(\l,\mu)H(\mu,\l) = -\rho^2(\l)  \left( {\sin\epsilon\over\epsilon} \right)^2
\eeq
with $\epsilon=N(\l-\mu)\rho(\l)$.
This result confirms the universality of the two point correlation function in a case more general than \cite{rBrZe}.
From this expression of $H$ and from Appendix 4, we can also find all the connected correlation functions with more than two points, in the short distance regime.
The result is still identical to the one-matrix-model's one.

\smallskip
Now let us come to the new aspect of this model: the mixed two point function. We will see that it is always zero at the order we consider.
We have from \rf{eDCxyK} and \rf{epsinapp}:
$$K(\l,\t\l)=\sum_{x,x^*,y,y^*} {1\over 2\pi N}{1\over x-y}{1\over \sqrt{-{\d\l\over\d{x}}{\d{\t\l}\over\d{y}}}} \ee{N(\l\t\l-\int^x \t\l(\xi)\l'(\xi)\d\xi -\int^y\l(\xi)\t\l'(\xi)\d\xi)}$$
$K$ is proportionnal to $1/N$, and vanishes at large $N$, except when $x\to y$. 
The correlation function thus vanishes as $1/N^2$.

The only domain in which this correlation could be larger stands along the curve $\t\l=\t\l(\l)$ in the $(\l,\t\l)$ plane (i.e. when $x=y$).
Unfortunately, this curve lies outside the interesting region $\l\in [a,b]$ and $\t\l\in [\t{a},\t{b}]$.
So, we find that in the large $N$ limit, the two matrices are uncorrelated.

However, let us suppose that: $x-y\ll 1$. Then:
$$ K(\l(x),\t\l(y))\sim {1\over 2\pi N(x-y) \sqrt{-\l'(x)\t\l'(x)}} \ee{-N{(x-y)^2\over 2}\t\l'(x)\l'(x)}$$
if we set $\epsilon=\sqrt{N}(x-y)\sqrt{-\l'(x)\t\l'(x)}$ we obtain:
$$ K\sim \Re {1\over \pi\sqrt{N}} {1\over \epsilon}\ee{\epsilon^2/2}$$
$$ \t{K}\sim \Re {1\over \pi\sqrt{N}}{1\over \epsilon}\ee{-\epsilon^2/2}$$
$$\l-\l(\t\l)={\epsilon\over\sqrt{N}}\sqrt{-{\d\l\over\d{\t\l}}}$$
$$ \rho^{(c)}_{1,1}(\l,\t\l)\sim {1\over N\pi^2\epsilon^2}$$
We still observe some kind of universality, but the meaning is not clear, since
it does not concern the true physical eigenvalues.

\subsection{ Smoothed two point functions}

When $\l-\mu$ is not small, as for one matrix, the correlation functions are not universal.
However, we see that equations \rf{epsinapp}\ contain very fast oscillations of frequency $N$.
At large distances, these oscillations can't be observed, the physical quantities are the smoothed functions.
So, inserting expressions \rf{epsinapp}\ in \rf{edefKer}\ and then in \rf{erhodeuKer}\ , we  find the complete two point correlation functions in the large $N$ limit.
Then we smoothing the large frequency oscillations, i.e. suppressing the terms of the form $\exp{N\int\t\l \l'}$, we get:
\beq\label{erhodeusmooth}
\begin{array}{rl}
\rho^{(c)}_{(2,0)}(\l,\mu)     & = {-1\over \pi^2N^2} {1\over 4} \sum_{\l(x)=\l,\l(y)=\mu} \, {1\over \l'(x)\l'(y)} {1\over (x-y)^2}      \cr
\rho^{(c)}_{(0,2)}(\t\l,\t\mu) & = {-1\over \pi^2N^2} {1\over 4} \sum_{\t\l(x)=\t\l,\t\l(y)=\t\mu} \, {1\over \t\l'(x)\t\l'(y)} {1\over (x-y)^2}
                                                                         \cr
\rho^{(c)}_{(1,1)}(\l,\t\l)    & = {-1\over \pi^2N^2} {1\over 4} \sum_{\l(x)=\l,\t\l(y)=\t\l} \, {1\over \l'(x)\t\l'(y)} {1\over (x-y)^2} \cr
\end{array}
\eeq
As in section (1.3), such smoothed correlations could also be obtained by the loop correlators.
The authors of \cite{rDKK}\ have calculated explicitely the two-loop correlators and found:
\beq\label{eomdeu}
\begin{array}{rl}
\om_{2,0}(z_1,z_2)         & = -\partial_{z_1}\partial_{z_2} \ln{x(z_1)-x(z_2)\over z_1-z_2}                                         \cr
\om_{0,2}(\t{z}_1,\t{z}_2) & = -\partial_{\t{z}_1}\partial_{\t{z}_2} \ln{x(\t{z}_1)-x(\t{z}_2)\over \t{z}_1-\t{z}_2}                         \cr
\om_{1,1}(z_1,\t{z}_2)     & = -\partial_{z_1}\partial_{\t{z}_2} \ln{\left( 1-{x(z_1)\over x(\t{z}_2)}\right)}                                      \cr
\end{array}
\eeq
Where the functions $x(z)$ and $x(\t{z})$ are defined by inverting \rf{elx}.
Both results coincide.

\medskip
As for one matrix, the other smoothed correlation functions can't be obtained by this method.
Indeed, since the kernels $K,H,\t{K},\t{H}$ are of order $1/N$, we can compute the $n$-point function only at order $1/N^n$, while we have seen by a topological argument that the $n$-loop function is of order $1/N^{2n-2}$.
To continue, it would be necessary to generalize the methods of \cite{rAJM}\ to the multi-matrix model.
However, this has not been done yet, but we can guess that the higher order
smoothed correlation functions can be expressed in terms of the functions $x(\l)$, and an increasing number of coefficients depending on the potentials $V$ and $\t{V}$.
They should present more universality than the 1-point function.
\subsection{Conclusions}

By the analysis and the exact calculation of the correlation functions of the 2-matrix model, we have confirmed with more general hypothesis the universality properties that Brezin and Zee have established for a 1-matrix model.
When one considers only one of the matrices, every thing seems to happen as in the 1-matrix case: the short distance 2-point correlation function \rf{erhodeumicro}\ is exactly the same.
But the long distance two point function is no more universal (cf \cite{rIsoKav}): it can be expressed in terms of an auxilary function $x(\l)$, but in the general case, it is not as simple as \rf{erhodeu}\ .
The functions $x(\l)$ and $x(\t\l)$ themselves, are not universal, they depend on the coefficient $\alpha,\alpha_k,\t\alpha_k$ (one of them can be eliminated by \eq{ealphasq}\ ), i.e. ${\rm deg} V+{\rm deg} \t{V}$ coefficients.
In other words we have just performed a convenient change of variables among the parameters $g_k,\t{g}_k$.

\smallskip
The new aspect of the two matrix model, is that we can check the correlation
between two coupled random matrices.
We then find that in the large $N$ limit, they are uncorrelated.
Indeed, the correlation function is most of time of order $1/N^2$.
The only domain in which this correlation could be larger is not physical, i.e.
when $\l$ and $\t\l$ verify $\t\l=\t\l(\l)$ (we have seen that if $\l\in [a,b]$, $\t\l(\l)$ is not real, indeed $\Im \t\l(\l)=\pi\rho(\l)$).

\smallskip
We have not treated the case of a chain of random matrices coupled by a weight of the form $\ee{N\tr M_k M_{k+1}}$, but it seems reasonable to think that
all the method developed in this article should apply with no difficulties to it , and it will be done in a following paper \cite{rCorreln}.
We would then find the same general feature: the short distance correlation function of one of the matrices is still \rf{erhodeumicro}\ .
The smoothed functions still look like \rf{erhodeusmooth}\ , but with more complicated functions $x(\l)$, involving a set of coefficient $\alpha_k$ equivalent to the set of the coefficient of the potentials.

\vfill\eject
%----------------------------------------------------------------%
\appendix{1}{Orthogonal polynomials $P_n$: an explicit expression.}

Let us show that the orthogonal polynomials $P_n$ defined by the orthogonality
condition: 
$$ \left<P_n \cdot P_m\right> = \delta_{nm} = \int \d \l\, \ee{-N V(\l)}\,
P_n(\l)P_m(\l) .$$ 
are given, up to a normalization, by equation \rf{ePnexact}:
$$ P_n(\l)\propto \int d^{n^2} M\, \ee{-N \tr V(M)} \det (\l -M).$$
First, this integral clearly yields a polynomial of degree $n$ in $\l$.
Let us then verify the orthogonality property:
after integration over the unitary group, the integration measure $\int \d M$
reduces to an integration over the eigenvalues of $M$, and the Jacobian of
this transformation is a square Vandermonde determinant: 
$$ P_n(\l) \propto\int \prod_{i=1,\ldots,n} \d \l_i\, \ee{-N
V(\l_i)}\left(\lambda-\lambda_i\right)  \,
\Delta^2(\l_1,\ldots,\l_n),$$ 
where 
$$\Delta(\l_1,\dots \l_n)=\prod_{i<j} (\l_i-\l_j).$$
In this form we recognize a more classical expression \cite{rSze}. Then, 
setting $\l=\l_0$: 
$$ <P_n(\l_0) \cdot \l_0^m>\propto\int \prod_{i=0,\ldots, n} \d\l_i\, \ee{-N
V(\l_i)} \, \Delta(\l_0,\l_1,\ldots ,\l_n)\, \Delta(\l_1,\ldots ,\l_n) \l_0^m.
$$   
The first Vandermonde is completely antisymmetric in the $n+1$ variables, we
can therefore antisymmetrize 
the factor $\Delta(\l_1,\ldots ,\l_n) \l_0^m$, the result is zero if $m<n$
because the only polynomial completely antisymmetric and of degree less than
$n-1$ in $\l_0$ is zero. Thus  
$$ <P_n \cdot \l^m> =0 \,.$$
it proves that $P_n$ is orthogonal to any $P_m$ with $m<n$, and that $<P_n.P_m>=0$ as
soon as $m\neq n$.

Remark that a similar integral representation can also be found for
multi-matrix models. Consider two families of orthogonal polynomials $P_n$ and
$\t{P}_n$, such that: 
$$<P_n(\l) \cdot \t{P}_m(\t\l)>=\int\int \d\l\,\d{\t\l}\, \ee{-N
(V(\l)+\t{V}(\t\l)-c\l\t\l)} P_n(\l) \t{P}_m(\t\l)\propto\delta_{nm}. $$ 
Then, we have:
$$P_n(\l)=\int\int \d{M} \, \d{\t{M}}\, \ee{-N\tr[ V(M)+\t{V}(\t{M})-cM\t{M}]}
\det(\l-M)$$  
where $M$ and $\t{M}$ are hermitian $n\times n$ matrices.
Similarly:
$$\t{P}_m(\t\l)=\int\int \d{M} \, \d{\t{M}}\, \ee{-N\tr[V(M)+\t{V}(\t{M})-cM\t{M}]}
\det(\t\l-\t{M}).$$

%----------------------------------------------------------------%
\appendix{2}{Connected correlation functions and the kernel
$\kappa(\l,\mu)$.} 

Some exact expressions exist for the correlation functions of matrix models \cite{rDyson},
in terms of a kernel $\kappa(\l,\mu)$.
We again consider the matrix distribution \rf{eprob}:
$$  {\cal P}(M)={1\over Z} \ee{-N\tr V(M)}.$$
The corresponding measure can be rewritten in terms of the eigenvalues
$\lambda_i$ of $M$ and a unitary transformation $U$ which diagonalizes $M$:
$$  {\cal P}(M)\d M=Z^{-1}\d U \prod_{i=1\ldots N} \d \l_i\, \ee{-N V(\l_i)}
\, \Delta^2(\l_1, \ldots,\l_N)$$
($\Delta=\prod_{i<j} (\l_i-\l_j)$ being the Vandermonde determinant).
Therefore, the probability that the eigenvalues of $M$ are $\l_1,\dots,\l_N$
is:
$$\rho_N(\l_1,\ldots ,\l_N)\propto \prod_{i=1\ldots N} \ee{-N V(\l_i)} \,
\Delta^2(\l_1,\ldots ,\l_N),$$ 
and the correlation functions are obtained by partially integrating over some
eigenvalues:
$$\begin{array}{rl}
\rho(\l_1)         & = \int \prod_{i=2,\ldots, N} \d \l_i\, \rho_N(\l_1,\ldots,\l_N) ,         \cr
\rho_2(\l_1,\l_2)  & = \int \prod_{i=3,\ldots, N} \d \l_i\,
\rho_N(\l_1,\ldots ,\l_N)          \cr
\end{array}$$ 
... and so on.

The Vandermonde determinant $\Delta$ can be written as:
$$ \Delta(\l_i)=\det \l_i^{j-1},$$
and thus, after some linear combinations of columns of the matrix:
$$ \Delta=\det \Pi_{j-1}(\l_i),\qquad
\Pi_n(\lambda)=\lambda^n+O\left(\lambda^{n-1}\right),  $$
identity true for any set of polynomials $\{ \Pi_n \}$ normalized as above. In order
to
perform the $\lambda$ integrations, we choose $\Pi_n\propto P_n$, $P_n$
being the orthogonal polynomials \rf{ePns}. $\Delta^2$ is the product of two such
determinants, therefore it is the determinant of a matrix product:
\bea \label{eKDel}
\Delta^2     & \propto \det K(\l_i,\l_j)                \label{eKDela}  \cr
K(\l_i,\l_j) & = \sum_{k=0}^{N-1} P_k(\l_i) P_k(\l_j).  \label{eKDelb}  \cr
\eea

The proportionality constant in  \eq{eKDela} is here irrelevant because the
eigenvalue distribution is normalized.
The Darboux--Christoffel formula (appendix 3) tells that: 
$$K(\l,\mu)=\alpha{P_N(\l)P_{N-1}(\mu)-P_N(\mu)P_{N-1}(\l)\over \l-\mu}$$
where $\alpha$ is a normalization constant depending on $N$, and
$\alpha=(a-b)/4$ when $N\to\infty$. 
The important properties of $K(\l,\mu)$ are:
$$\begin{array}{rl}
\int\d\nu\, \ee{-NV(\nu)} K(\l,\nu)           & = 1\,,\quad
\int\d\nu\, \ee{-NV(\nu)} K(\nu,\nu) = N\,,                       \cr
\int\d\nu\, \ee{-NV(\nu)} K(\l,\nu)K(\nu,\mu) & = K(\l,\mu).      \cr
\end{array}$$ 
An explicit expression for $\rho_n$ 
$$\rho_n(\l_1,\dots,\l_n)={1\over N!}\int \prod_{i=n+1,\dots, N} \d\l_i\,
\prod_{i=1,\ldots, N}\ee{-N V(\l_i)} \,\det  K(\l_i,\l_j),$$ 
can be obtained by successively integrating over eigenvalues.
Using the rules of $K$ integration it is easy to prove by induction
$$\rho_n\left(\lambda_1,\lambda_2,\ldots,\lambda_n\right)=
{N^n(N-n)!\over N!}\det \kappa\left(\lambda_i,\lambda_j\right),$$
where we have introduced the reduced function
$$\kappa(\lambda,\mu)={1\over
N}\ee{-(N/2)[V(\lambda)+V(\mu)]}K(\lambda,\mu)\,.$$ 
Therefore, when the $\l_i$ are all distinct we have:
$$\left< \O(\l_1)\ldots \O(\l_n)\right>=\det
\kappa(\l_i,\l_j)=\sum_\sigma (-1)^\sigma \prod_i \kappa(\l_i,\l_{\sigma_i})$$
The connected function will involve only the sum over the permutations $\sigma$
such that $\prod_i \kappa(\l_i,\l_{\sigma_i})$ cannot be split in the product
of two cyclic products, i.e. only cyclic permutations will contribute to the
connected function. This intuitive result is a classical combinatorial identity. %It
can be proven from% 
%a simple identity (obtained for instance by fermion integration%
%methods),%
%$$\det \left[\delta_{ij}+s_i\kappa(\lambda_i,\lambda_j)\right]=%
%\sum_n{1\over n!}\sum_{i_1,i_2,\ldots,i_n}s_{i_1}s_{i_2}\ldots%
%s_{i_n}\det{}^{(n)}\kappa\left(\lambda_{i_l}\lambda_{i_k}\right),$$%
%Taking the logarithm of both sizes, and expanding in power series of the% %sources%
%$s_i$, one obtains %
The connected function can thus be written:
$$\begin{array}{rl}
\left< \O(\l_1)\ldots \O(\l_n) \right>_{\rm conn}   & = (-1)^{n+1} {1\over n} \sum_{{\rm permutations}\ \sigma} \prod_i \kappa(\lambda_{\sigma_i},\l_{\sigma_{i+1}}),                \cr
                                                    & = (-1)^{n+1} \left[\kappa\left(\lambda_1,\lambda_2\right)
\kappa\left(\lambda_2,\lambda_3\right) \ldots
\kappa\left(\lambda_{n},\lambda_1\right) + \cdots \right] ,  \cr
\end{array}$$
where the additional terms in the r.h.s.\ symmetrize the expression.
Or in a more compact way:
$$ \left< \O(\l_1)\ldots \O(\l_n)\right>_{\rm
conn}=(-1)^{n+1}
\sum_{{\rm cyclic\, permutations}\
\sigma\,}\,\prod_{i=1}^n \kappa\left(\l_i,\lambda_{\sigma_i}\right)$$
%--------------------------------------------------------%

\appendix{3}{Derivation of the Darboux--Christoffel formula.}

The polynomial $\l P_n(\l)$ can be expanded on the basis of the $P_m$ with
$m\leq n+1$: 
$$\l P_n(\l)= \sum_{m=n-1}^{n+1}Q_{nm} P_m(\l).$$
The orthogonality condition \rf{ePns}\ implies that the matrix $Q$ is symmetric:
$$Q_{nm}=<\l P_n \cdot P_m>=<P_n \cdot \l P_m>=Q_{mn}\,.$$
The polynomial $(\l-\mu)K(\l,\mu)$ can thus be written:
$$(\l-\mu)K(\l,\mu)=\sum_{k=0}^{N-1}\left(
\sum_{i=0}^N Q_{ki}P_i(\l)P_k(\mu)-\sum_{j=0}^N Q_{kj}P_k(\l)P_j(\mu) \right).$$
All the terms cancel, except the upper-bounds: 
$$(\l-\mu)K(\l,\mu)=Q_{N,N-1} \bigl( P_N(\l)P_{N-1}(\mu)-P_N(\mu)P_{N-1}(\l)
\bigr),$$ 
therefore $\alpha=Q_{N,N-1}$.
In the large $N$ limit, it is possible to calculate $\alpha$. The simplest way
of doing this is to calculate $\l P_{N-1}$ from expression \rf{ePnasym}. Since
$$\left( {a+b\over2}-{b-a\over2}\cos\varphi\right)\cos{(\psi-\varphi)}=
{a+b\over2}\cos{(\psi-\varphi)}-{b-a\over 4}\left(\cos\psi+\cos{(\psi-2\varphi)}\right)
,$$
we have:
$$\l P_{N-1}={a+b\over 2}P_{N-1}-{b-a\over4}\left( P_N+P_{N-2}\right)$$
and therefore $\alpha=(a-b)/4$.
%--------------------------------------------------------------------------
%\bigskip
\appendix{4}{Partially integrated distributions of the 2-matrix-model.}
\medskip

Exact expressions for the correlation functions have been recently derived in \cite{rMehtaker}\ . Hereafter, we just give the results and present a diagrammatical method to express the partially integrated distributions of type \rf{eronmpid} in terms of the kernels $H,K,\t{H},\t{K}$.
For instance, generalizing the one-matrix result, we have:
$$-{\rho_{n,0}}^{(c)}(\l_1,\dots,\l_n)=(-1)^{n} {1\over n} \sum_{\sigma} \prod_{i=1}^n H(\l_{\sigma_i},\l_{\sigma_{i+1}})$$
We observe that $\rho_{n,o}^{(c)}$ is a symetric sum of cyclic products of kernels $H$. The general $\rho_{n,m}^{(c)}$ have the same form, except that the other kernels appear. The method consists in writting all the possible permutations of the eigenvalues $\l_1,\dots,\l_n,\t\l_1,\dots,\t\l_m$, and multiply the kernels of two consecutive eigenvalues in a cyclic product. All this can be represented in a diagramatic method:

-First write a permutation of all the $n+m$ eigenvalues as a closed oriented chain.

-assign to each link between two consecutive eigenvalues a factor according to the following rules:
$$\begin{array}{cc}
     \l_i\longrightarrow\l_j \, = -H(\l_i,\l_j) \qquad
&    \qquad \l_i\longrightarrow\t\l_j \, = -K(\l_i,\t\l_j)               \cr
     \t\l_i\longrightarrow\t\l_j \, = -\t{H}(\t\l_i,\t\l_j) \qquad
&    \qquad \t\l_i\longrightarrow\l_j \, = {1\over N}-\t{K}(\t\l_i,\l_j) \cr
\end{array}$$

-then $-\rho^{(c)}$ is the sum for all these diagramms of the products of the links.  You can verify the number of terms in the sum is
$$ \sum_{k=0}^{{\rm min}(n,m)} {n!\over n-k!}{m!\over m-k!}(n+m-k-1)!$$

For instance we have already given the one and two point functions in \rf{erhodeuKer}, we write also here the 3-point functions:
$$\begin{array}{rl}
\rho_{(2,1)}^{(c)} (\l_1,\l_2,\t\l) =
& \t{K}(\t\l,\l_1) H(\l_1,\l_2) K(\l_2,\t\l) + \t{K}(\t\l,\l_2) H(\l_2,\l_1) K(\l_2,\t\l)                   \cr
 & -{1\over N} H(\l_1,\l_2) K(\l_2,\t\l) - {1\over N} H(\l_2,\l_1) K(\l_1,\t\l)
                               \cr
\end{array}$$

%
%--------------------------------------------------------------------------
\appendix{5}{An approximation for the two matrices orthogonal polynomials.}
\medskip

The proof follows the same method as in the one-matrix case:
we have the integral representation (Appendix 1)
$$ P_n(\l)={\rm cte} \int dM^{n\times n} d\t{M}^{n\times n} \ee{-N\tr \left( V(M)+\t{V}(\t{M})-M\t{M}\right)} \det (\l-M)$$
wich we write:
$$ P_n(\l)=Z\left(\l,g={n\over N},h={1\over N}\right)=\ee{n^2 F(\l,g,h)}$$
$$ Z(\l,g,h)= \int \d{M^{n\times n}}\d{\t{M}^{n\times n}} \ee{-{n\over g}\tr \left( {\cal V}(M)+\t{V}(\t{M})-M\t{M}\right)}$$
$$ {\cal V}(z)=V(z)-h\ln{\l-z}$$
We observe that
$$ n{d F(\l,g,h)\over d\l}=\om(z=\l,\l,g,h)$$
where $\om(z)$ is the resolvent corresponding to $Z(\l,g,h)$.
We need the leading behaviour as $N\to\infty$ and $N-n\sim1$ of $n^2F$, and we need only its dependance in $\l$.
$$ n^2{ d F\over d\l}=Ng\om$$
We  have to expand $\om$ in the limit $g\to 1$ and $h\to 0$:
$$g\om=\om_0+(g-1)\Om_g+h\Om_h+ O({1\over N^2})$$
with
$$\om_0(z)=\om(z,g=1,h=0) ,\qquad
{\begin{array}{rl}
 \Om_g  & = \left. {\partial g\om\over \partial g}\right|_{g=1,h=0}   \cr
 \Om_h  & = \left. {\partial g\om\over \partial h}\right|_{g=1,h=0}   \cr
\end{array}}$$
Then:
\beq\label{edFsurdl}
 n^2{ d F\over d\l}=N\om_0(\l)-(N-n)\Om_g+\Om_h +O(1/N^2)
\eeq
We have already noted that:
$$\om_0(z)=V'(z)-\t\l(z)$$
and we can determine $\Om_{g,h}$ by the remark that a variation of the potential by an amount of $\delta V(z)$, induces a variation of $\om$ given by the two-point functions $\om_2={\partial \om /\partial V}$:
\beq\label{evaromV}
 \delta \om(z_1)=-{1\over 2i\pi}\oint \d{z_2} \om_{2,0}(z_1,z_2) \delta V(z_2)+\om_{1,1}(z_1,z_2)\delta\t{V}(z_2)
\eeq
$\Om_g$ corresponds to the variation of the potential $\delta V=1/g V-V\sim -(g-1)V$, and $\Om_h$ to $\delta V= -h\ln{\l-z}$.
Inserting the explicit explicit expressions \rf{eomdeu} of \cite{rDKK}\ into \rf{evaromV} , and integrating by parts and taking the residues, we obtain:
$$\Om_h(z,\l)=\partial_z \left(\ln{x(z)-x(\l)\over z-\l}\right)$$
at $z=\l$ we have:
$$\Om_h(\l)={1\over2}{d\over d\l} \ln{d x(\l)\over d\l}$$
Remark that for $\Om_g$ there is a simpler method: we have $\psi_n=\hat{x}^{n-N}\psi_N$, the $(N-n)$ dependance is thus $x^{n-N}$.
Therefore, we know that 
$$\Om_g(\l)={d\over d\l} \ln{x(\l)}$$

Inserting these expressions in \rf{edFsurdl} and integrating with respect to $\l$, we obtain the asymptotic expression of $\psi_n(\l)$:
\beq\label{epsinapprox}
 \psi_n(\l)\sim {\rm cte} \sqrt{d x\over d\l} \ee{-N\int_{x_0}^x \t\l(\l)\d\l } x^{n-N}
\eeq
Unfortunately, this result is obviously wrong.
Let us remember that in the one-matrix case, at this point, we had considered the real part of this expression, and replaced the exponential by a $\cos{}$. Another way to understand what happened, is that in the classical limit, we have replaced operators by numbers.
A more carefull analysis would have given instead of \rf{edifpsin}\ a differential linear equation of order ${\rm deg} \t{V}$ for $\psi_n$, and thus we should add independent solutions which all have the form \rf{epsinapprox}, but with the different values of $x$ satisfying \eq{elx}\ (Thanks to F. David for this remark).
\beq\label{epsinappp}
 \psi_n(\l)\sim \sum_{\l(x_k)=\l}\, {\rm cte}_k \,\sqrt{dx_k\over d\l} \ee{-N\int_{x_0}^{x_k} \t\l(y)\l'(y) \d{y}} x_k^{n-N}
\eeq
The function $x(\l)$ is defined as the physical one, i.e. such that $V'(\l)-\t\l(x)=\om(\l)$ is the resolvent and behaves as $1/z$ for $z\to\infty$.
It is the one such that $\l(x)\mathop\sim_{x\to\infty} \alpha x$.
At large $\l$, only this solution contributes to \rf{epsinapprox}\ , while along the cut $\l\in [a,b]$, all the $x$ contribute.
We then have:
\beq\label{epsinapppp}
 \psi_n(\l)\sim {\rm cte}\, 2\sum_{x} \, \sqrt{dx\over d\l} \ee{-N\int_{x_0}^{x} \t\l(y)\l'(y) \d{y}} x^{n-N} .
\eeq

The same method gives also $\t\psi_n$, and to obtain the $\chi_n$ you just integrate \rf{edefchi}\ by steepest descent.
Note that the ${\rm cte}$ factor can be fixed by the normalizations.
At the end, you find \rf{epsinapp}\ .

\appendix{6}{The loop equations for the two matrix model.}
\medskip

The loop equation reflect the invariance by a change of variable into the partition function:
$$ Z= \int \,\d M \,\d {\t{M}} \,\,\ee{-N\tr (V(M)+\t{V}(\t{M})-M\t{M})} $$
First consider the change of variable:
$$ M\to M+ \epsilon {1\over z-M}{1\over \t{z}-\t{M}} $$
At leading order in $1/N$ (i.e. approximating $<Tr Tr>$ by $<Tr><Tr>$) we get :
$$ \left(V'(z)-\om(z)-\t{z} \right) \om(z,\t{z})= \sum_{i,j} p_{ij} z^i {\t\om}_j(\t{z}) \,\, -\om(z) $$
where we have defined
$$ \om(z,\t{z})=\moy{ \Tr {1\over z-M}{1\over \t{z}-\t{M}} } \qquad , \qquad \om_j(z)=\moy{ \Tr {1\over z-M}\t{M}^j }$$
$${\rm and}\quad P(x,y)=\sum_{i,j =0}^{{\rm deg}V'-1} \, p_{ij}x^i y^j = {V'(x)-V'(y)\over x-y} $$
Making the symetrical change of variable for $\t{M}$ we find that:
$$ \om(z,\t{z}) = {\sum_{i,j} p_{ij} z^i {\t\om}_j(\t{z}) \,\, -\om(z) \over  V'(z)-\om(z)-\t{z}} = {\sum_{i,j} \t{p}_{ij} \t{z}^i \om_j(z) \,\, -\t\om(\t{z}) \over  \t{V}'(\t{z})-\t\om(\t{z})-z} $$
then reducing to the same denominator, we get that
$$ \left(  \t{V}'(\t{z})-\t\om(\t{z})-z \right)\sum_{ij} p_{ij} z^i{\t\om}_j(\t{z}) \,\, -\t{z}\t\om(\t{z}) +V'(z)\t\om(\t{z}) = {\rm idem\, with}\,\, z\leftrightarrow \t{z} $$
The left hand side is a polynomial in $z$ of degree deg$V'-1$, while the right hand side is a polynomial in $\t{z}$ of degree deg$\t{V}'-1$.
Therefore Both sides are polynomials, let us call this polynomial $Q(z,\t{z})$.
This allows to write some algebraic equation for $\om(z)$.
Moreover, we have an expression of $\om(z,\t{z})$:
$$ \om(z,\t{z}) = 1+ {Q(z,\t{z}) - (V'(z)-\t{z})(\t{V}'(\t{z})-z) \over  (V'(z)-\om(z)-\t{z})(\t{V}'(\t{z})-\t\om(\t{z})-z)} $$
The poles must cancel, so:
$$ Q(z,\t\l(z)) = (V'(z)-\t\l(z))(\t{V}'(\t\l(z))-z) \qquad {\rm when} \quad \t{z}=\t\l(z)=V'(z)-\om(z) $$
$$ Q(\l(\t{z}),\t{z}) = (V'(\l(\t{z}))-\t{z})(\t{V}'(\t{z})-\l(\t{z})) \qquad {\rm when} \quad z=\l(\t{z})=\t{V}'(\t{z})-\t\om(\t{z}) $$
These algebraic equations can be shown to be equivalent to the equations of motion \rf{eqmotioncl}\ found by the method of orthogonal polynomials.

\vskip 4cm
%\listrefs

\bibliography{abbrv}
\end{document}